\documentclass[10pt,twocolumn,letterpaper]{article}

\usepackage{wacv}              %

\usepackage{multirow}
\usepackage[accsupp]{axessibility}  %
\usepackage[dvipsnames]{xcolor}

\newif\ifdraft
\drafttrue
\ifdraft
\newcommand{\smc}[1]{{\color{blue}[\textbf{SM:} #1]}}

\else
\newcommand{\smc}[1]{}

\fi

\newcommand{\methodname}{\text{Snapmoji}}
\newcommand{\lego}{\text{LEGO}}
\newcommand{\yoda}{\text{Yoda}}

\definecolor{wacvblue}{rgb}{0.21,0.49,0.74}
\usepackage[pagebackref,breaklinks,colorlinks,allcolors=wacvblue]{hyperref}

\def\wacvPaperID{2576} %
\def\confName{WACV}
\def\confYear{2026}

\title{Snapmoji: Instant Generation of Animatable Dual-Stylized Avatars}

\author{Eric Ming Chen$^{1,2*}$ \quad Di Liu$^{1,3*}$ \quad Sizhuo Ma$^1$ \quad Michael Vasilkovsky$^1$ \quad
Bing Zhou$^1$ \quad Qiang Gao$^1$ \\
Wenzhou Wang$^1$ \quad Jiahao Luo$^{1,4}$ \quad
Dimitris N. Metaxas$^3$ \quad Vincent Sitzmann$^2$ \quad Jian Wang$^1$ \vspace{+0.3em} \\
$^1$Snap Inc. \quad $^2$MIT \quad
$^3$Rutgers University \quad $^4$University of California, Santa Cruz \vspace{-0em} \\
}

\begin{document}

\twocolumn[{%
\renewcommand\twocolumn[1][]{#1}%
\maketitle
\includegraphics[width=\linewidth]{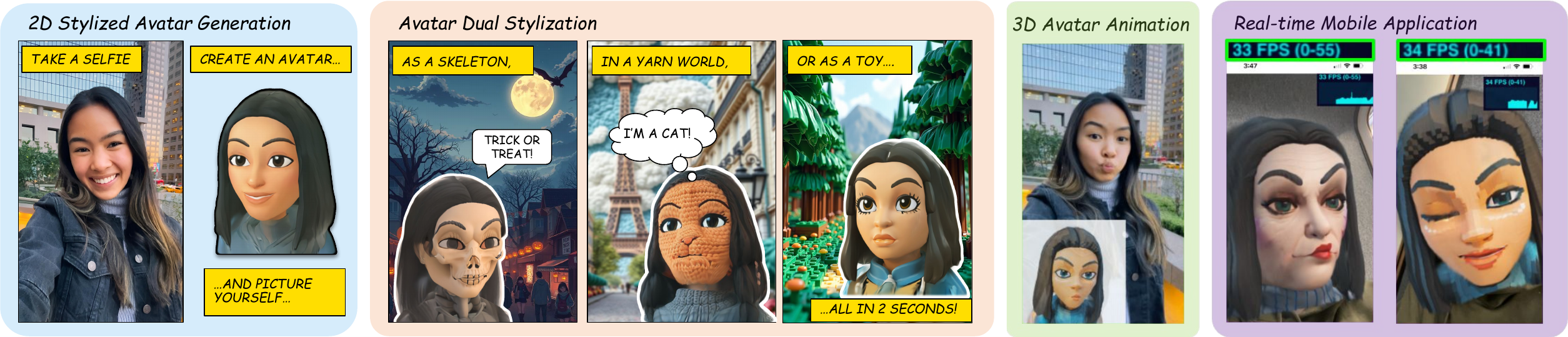}
\vspace{-10pt}
\captionof{figure}{We introduce \textbf{\methodname{}}, a system that can instantly generate animatable dual-stylized avatars. Our dual stylization process reimagines a base avatar in various artistic styles, enabling users to visualize themselves in diverse scenarios and create personalized stories. \methodname{} accomplishes the selfie-to-avatar conversion in just 0.9 seconds. Equipped with a mobile face tracker, our approach also enables expression animation in real time, at 30 to 40 FPS on mobile devices. \href{https://echen01.github.io/instamoji-supp/}{Project page.}\vspace{2em} }
\label{fig:teaser}
}]

\makeatletter
\renewcommand\@makefntext[1]{%
  \setlength{\parindent}{0pt} %
  \@makefnmark #1%
}
\makeatother
\def\thefootnote{*}\footnotetext[1]{Equal contribution}%
\begin{abstract}
Despite the increasing popularity of avatar systems such as Snapchat Bitmojis, existing production avatar platforms face several limitations, such as a limited number of predefined assets, tedious customization processes, and inefficient rendering requirements. Addressing these shortcomings, we introduce \methodname{}, an avatar generation system that instantly creates 3D avatars, and enables customization in a process we call dual-stylization. \methodname{} first maps a selfie of a user to a primary avatar (e.g., Bitmoji style) using a new technique we name Gaussian Domain Adaptation (GDA), then applies a secondary style (e.g., skeleton, yarn, toy) to the primary avatar, all while preserving the user’s identity. The generated 3D avatars can then be rendered an animated on mobile devices at 30--40 FPS. %
\end{abstract}

\section{Introduction}
\label{sec:intro}

 With platforms like Snapchat Bitmojis~\cite{Bitmoji}, Apple Memojis~\cite{Memoji}, and Meta avatars~\cite{Meta_avatar}, personalized avatars have now become a cornerstone of social media. These platforms allow users to choose features from a range of cartoon-like traits, such as hairstyles, facial features, clothing, and accessories. However, despite their popularity, existing avatar platforms all share a central limitation: traits are limited to a list of predefined assets. Creating new traits requires a team of artists to create new assets from scratch, and this workflow becomes especially unsustainable as avatars are continuously updated. Take Snapchat for example, which hosts hundreds of avatar fashion assets, stickers, and animations. What happens if a user wants to picture their Bitmoji as if they were a superhero like Spiderman? On Snapchat, a typical marketing promotion for a movie or event will include 30 to 150 new clothing assets, requiring a significant amount of work from artists. So is there a way we can automatically stylize a user's avatar without this manual effort?
This challenge underscores the need for an ability we call \textbf{dual-stylization}: the ability of a platform to not only enerate a single avatar for a user, but also to re-style it across multiple themes without manually building new 3D assets. In this example, we call the user's base Bitmoji the \textit{primary avatar}, and their Bitmoji pictured as Spiderman their \textit{dual-stylized avatar}. 

Although recent research like StyleAvatar3D~\cite{zhang2023styleavatar3d}, TextToon~\cite{Song2024TextToonRT}, and DATID-3D~\cite{kim2022datid3d} propose generative methods to create 3D avatars without requiring new assets, we find that these methods still do not fulfill the requirements that a production system should have. Their generated avatars are typically expensive to create, cannot be rendered in real time, and are not animatable (see Table~\ref{tab:related_works} for a comparison). For applications like augmented reality (AR), they are unsuitable. 

\begin{table}[t]
\centering
\renewcommand\tabcolsep{4pt} %
\caption{Comparison among stylized avatar generation methods.}
\resizebox{\linewidth}{!}{ %
\begin{tabular}{c|c|c|c|c|c}
\toprule
Method & Selfie Input & Mobile AR& Asset-free & Animatable & Dual Style \\
\midrule
StyleAvatar3D~\citep{zhang2023styleavatar3d} & \phantom{0} & \phantom{0} & \checkmark & \phantom{0} & \phantom{0} \\
DATID-3D~\citep{kim2022datid3d} & \checkmark & \phantom{0} & \checkmark & \phantom{0} & \phantom{0} \\
TextToon~\citep{Song2024TextToonRT} & \phantom{0} & \phantom{0}  & \checkmark & \checkmark & \phantom{0} \\
EasyCraft~\citep{wang2025easycraftrobustefficientframework} & \phantom{0} & \checkmark & \phantom{0} & \checkmark & \phantom{0} \\
SwiftAvatar~\citep{Wang2023SwiftAvatarEA} & \checkmark & \checkmark & \phantom{0} & \checkmark & \phantom{0} \\

AgileAvatar~\citep{Sang2022AgileAvatarS3} & \checkmark & \checkmark & \phantom{0} & \checkmark & \phantom{0} \\

\methodname~(ours) & \checkmark & \checkmark & \checkmark & \checkmark & \checkmark \\
\bottomrule
\end{tabular}
}
\vspace{-15pt}
\label{tab:related_works}
\end{table}

In search for an alternative, we propose \textit{\methodname{}}, a novel framework for generating expressive and animatable avatars, represented in the form of 3D Gaussian Splats~\cite{Kerbl20233DGS}. Using data collected from the public APIs of Bitmoji~\cite{snapAPI}, \methodname{} is trained to map selfies of users to Bitmoji-styled avatars, and to support dual stylization: customizing the Bitmoji avatar in a user-specified style. The dual-styles are specified from text prompts, such as of ``\lego" or ``\yoda", and the generated avatars preserve both the primary Bitmoji style, and the user's identity.

Our system is designed with three core objectives in mind: 
\begin{enumerate}
\item \textbf{Dual Stylization:} The system should generate avatars in the Bitmoji art style, and a secondary style, such as of \lego{} or \yoda{}, while preserving user identity.
\item \textbf{User Convenience:} For ease of use, the system should require only a single image as input and produce the avatar instantly.
\item \textbf{Efficiency:} The avatars should enable real-time rendering on mobile devices, supporting applications like AR. 
\end{enumerate}

To meet these goals, we introduce a two-stage pipeline. In the first stage, we present Gaussian Domain Adaptation (GDA), a domain translation method that leverages a learned 3D prior to map realistic selfies into the Bitmoji space, followed by a diffusion model that further customizes the style based on user-provided text prompts. In the second stage, the stylized 2D avatar is lifted to a 3D avatar, and can be animated using blendshapes in real time.

Although our system is demonstrated using Bitmojis, the pipeline is general and can be adapted to other avatar ecosystems, potentially enabling a wide range of creative applications across gaming, social media, virtual meetings, and education.
In summary, our contributions are as follows:
\begin{itemize}[leftmargin=10pt]
\item We introduce the concept of dual stylization, and propose an efficient system to convert a realistic image into a dual-stylized 3D animatable avatar
\item  We propose a novel domain adaptation method, Gaussian Domain Adaptation (GDA), to transfer a real image into a predefined avatar style.
\item We develop a Javascript framework to animate our generated avatars on mobile phones at 30-40 FPS. 
\end{itemize}

\section{Related Work}

\begin{figure*}[t] 
\centering
\includegraphics[width=0.95\linewidth]{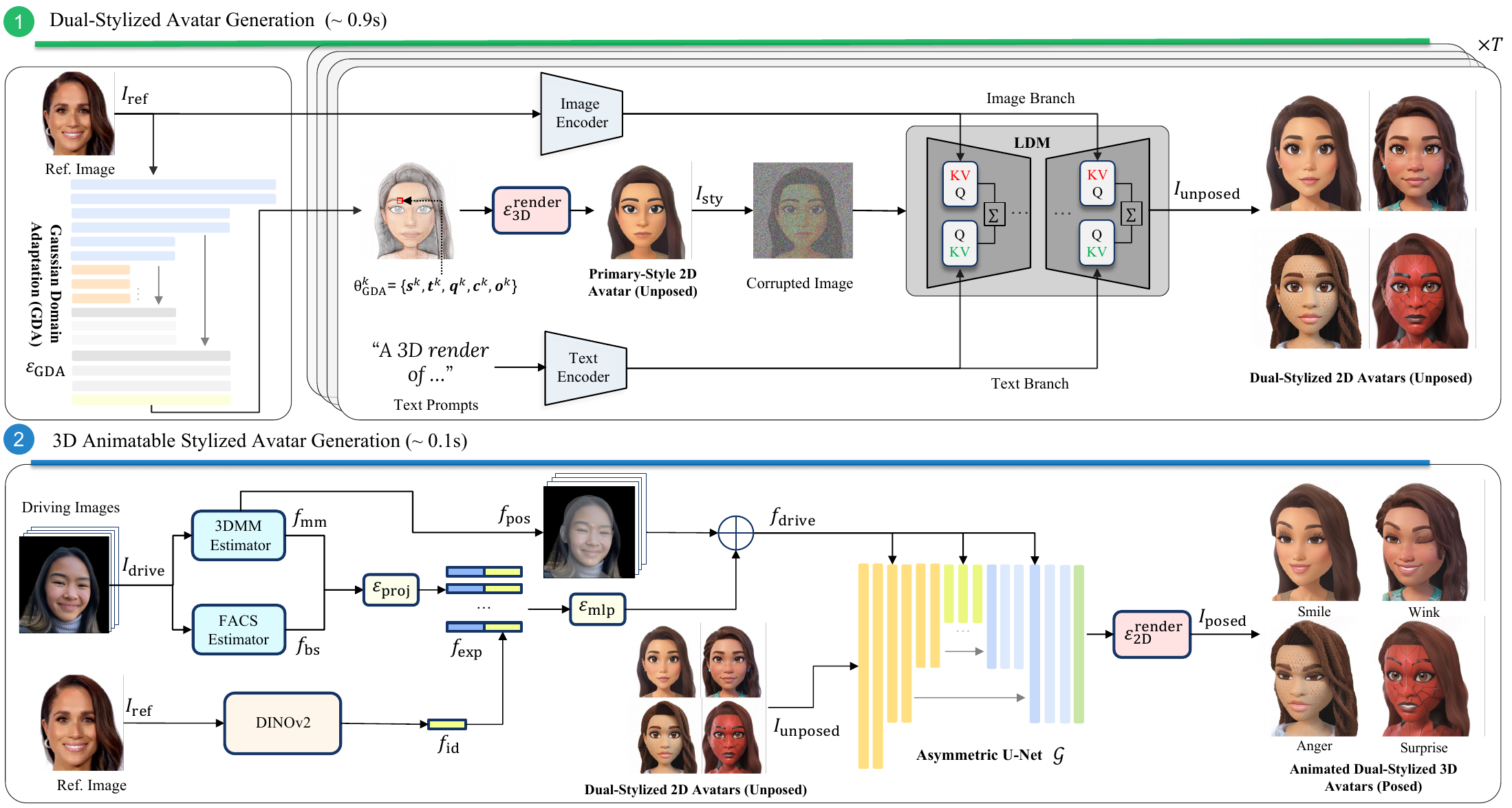} \caption{\textbf{The \methodname{} Inference Pipeline.} 
The pipeline has two stages. First, the Gaussian Domain Adaptation network $\mathcal{E}_\text{GDA}$ converts a facial image into a primary avatar $I_\text{sty}$. This avatar undergoes further personalization using a text-guided diffusion process with $T$ steps for additional stylization. Second, expression codes extracted via an 3DMM and FACS are combined with identity features $f_\text{id}$ from a reference image $I_\text{ref}$ and positional maps $f_\text{pos}$ from a driving image $I_\text{drive}$. The unposed dual-stylized avatar $I_\text{unposed}$ is then processed by an asymmetric UNet $\mathcal{G}(\cdot)$, conditioned on the driving codes $f_\text{drive}$ through cross-attention, to generate animated, dual-stylized 3D avatars.} \label{fig:img2avatar} 

\end{figure*}

\label{sec:related_work}

\paragraph{Commercial Systems for Avatar Creation.}
Commercial avatar platforms from TikTok, Apple, etc. have begun to introduce automated processes for creating avatars by training classifiers that can predict avatar traits from a user's photograph. The users can then manually adjust the traits to their preference. However, these classifiers require paired data that link real faces to specific avatar traits, which is difficult to acquire at scale. To address this challenge, approaches like AgileAvatar \citep{Sang2022AgileAvatarS3} and SwiftAvatar ~\citep{Wang2023SwiftAvatarEA} from TikTok; and  F2P~\citep{Shi2019FacetoParameterTF,Shi2020FastAR} and EasyCraft~\citep{wang2025easycraftrobustefficientframework} from NetEase; develop self-supervised learning techniques to map real photos to avatars. While these methods are efficient, they can only predict traits from existing 3D asset libraries. Our goal however is to generate entirely new avatars without requiring more 3D assets to be manually created. 

\paragraph{2D Stylized Avatar Generation} Also relevant to avatar creation are techniques in image-to-image translation. Neural networks like StyleGAN~\cite{Karras2019AnalyzingAI} have been used to map realistic images of faces to various domains, like of Disney cartoons, paintings, and vintage photos~\cite{Pinkney2020ResolutionDG,Chen2022WhatsIA,Luo-Rephotography-2021}. One reason why StyleGAN is so popular for this face stylization task is that it does not require paired images between the domains. Works such as SwiftAvatar~\cite{Wang2023SwiftAvatarEA} use this property of StyleGAN to map real faces to cartoon avatars. 

Along with GANs, diffusion models are another prominent approach for 2D stylization. Unlike GANs which are trained on single-class datasets, diffusion models are typically trained on internet-scale data, giving them more stylistic diversity than their GAN counterparts. To control the generation process, models like Stable Diffusion \cite{Rombach2021HighResolutionIS} take a text prompt as input. Further tools, including SDEdit \cite{Meng2021SDEditGI}, ControlNet \cite{Zhang2023AddingCC}, and IP Adapter \cite{Ye2023IPAdapterTC}, provide additional control by allowing image prompts as input. Because of their complimentary strengths, our method uses both GANs and diffusion models for avatar generation. 

\paragraph{3D Avatar and Object Generation.} Recent advances in 3D representations like neural radiance fields~\cite{mildenhall2020nerf} and Gaussian splats~\cite{Kerbl20233DGS} can also be applied to avatar generation. For instance, DATID-3D \cite{kim2022datid3d} and StyleAvatar3D \cite{zhang2023styleavatar3d} employ 3D GANs to generate and stylize 3D face models. More recent developments utilize text-to-image diffusion models for avatar stylization~\cite{Song2024TextToonRT,NguyenPhuoc2023AlteredAvatarSD,Men_2024_CVPR,he2024dice,dao2025improved,han2024proxedit}, though this process tends to be slow. 
Different from our task of cartoon avatar generation, much work also also been done in generating photorealistic avatars with Gaussian splats~\cite{saito2024rgca,qian2023gaussianavatars,ma2024gaussianblendshapes,liu2025lucaslayereduniversalcodec} . The major difference from these photorealistic avatar methods and cartoon avatar methods is that the former fits Gaussians on the surfaces of 3D Morphable Models (3DMMs) like FLAME \cite{FLAME:SiggraphAsia2017}. Because 3DMMs only model realistic human geometry, this strategy cannot be applied to cartoon avatars, which often have large geometric differences like big eyes or big heads.  

In addition to 3D avatars, our method is also inspired by work general 3D object generation. Notably, LRM~\cite{Hong2023LRMLR} and LGM~\cite{Tang2024LGMLM} are trained on Objaverse to map 2D images to 3D objects in a single network evaluation. As we show in Section~\ref{sec:gda}, this data prior from Objaverse can surprisingly be repurposed for image-to-avatar generation.

\section{Method}
Our method begins with creating a synthetic dataset of real face images and their 2D primary avatars (Sec. \ref{sec:dataset}), which is then used for GDA and dual stylization (Sec. \ref{sec:gda}). The dual-stylized avatars are then lifted to 3D and animated (Sec. \ref{sec:3d_gen}). Training details are provided in Sec. \ref{sec:training_losses}.

\subsection{Datasets}
\label{sec:dataset}

Training our image-to-avatar GDA model requires paired datasets of real faces and primary avatars, which are not available at scale. To overcome this, we use GAN inversion, similar to prior work in unsupervised domain adaptation \citep{Wang2023SwiftAvatarEA}, to create synthetic paired data. By aligning the latent spaces of a source GAN and a fine-tuned target GAN~\cite{Wu2021StyleAlignAA,Chen2022WhatsIA,Wang2023SwiftAvatarEA}, we generate corresponding pairs of realistic and Bitmoji faces. Specifically, Bitmoji images are inverted into the target GAN's latent space to obtain latent codes, which are then applied to the source GAN to produce realistic counterparts:
\begin{equation}
    w := \text{argmin}_{w \in \mathcal{W}} \| G_{\text{tgt}}(w) - I_{\text{tgt}} \|, \\
    I_{\text{src}} = G_{\text{src}}(w).
\end{equation}
Using this method, we generated 13,000 synthetic image pairs from Bitmoji avatars, forming the basis for GDA training. Examples are shown in the supplementary material. 

\subsection{Image to 2D Avatar Generation}
\label{sec:gda}

\noindent \textbf{Gaussian Domain Adaptation} $\mathcal{E}_{\text{GDA}}(\cdot)$\textbf{.} Our first step is to map real photos to their corresponding primary avatars (\ie, Bitmoji avatars). Surprisingly, we find that features learned by Large Multi-view Gaussian models (LGMs)~\cite{Tang2024LGMLM} can be readily adapted for style transfer. We believe this is due to their ability to hold internet-scale information from  multi-view training datasets such as Objaverse~\cite{Deitke2022ObjaverseAU}. We repurpose LGM for this style transfer task, and call this technique \textit{Gaussian Domain Adaptation}. In the supplementary material, we perform an ablation study demonstrating that GDA training struggles without Objaverse pre-training. 

\begin{figure}[t]
\centering
\includegraphics[width=1\linewidth]{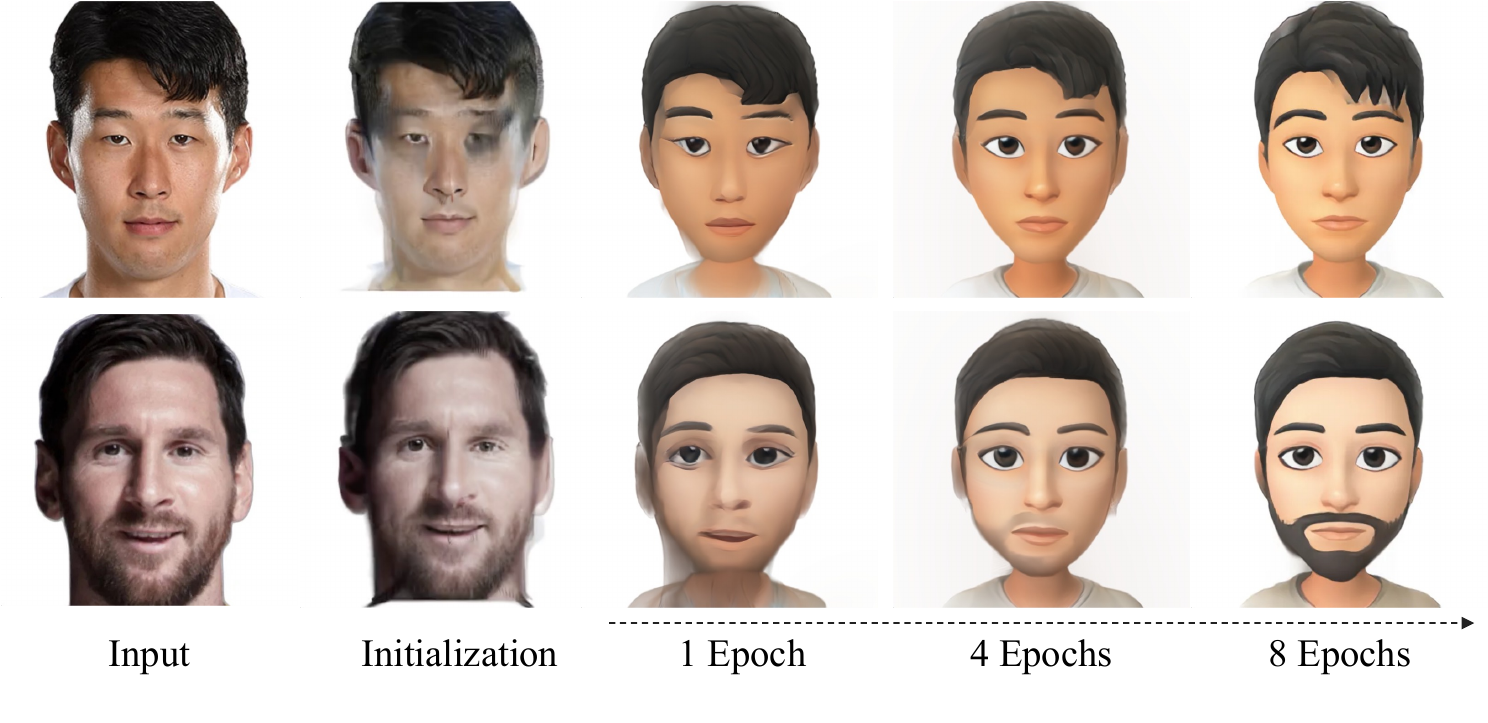}
\vspace{-15pt}
\caption{\textbf{Gaussian Domain Adaptation.} We show the outputs of the GDA network over several training epochs to visualize the domain shifts from natural images to cartoon avatars.}
\label{fig:domain}
\end{figure}

To adapt LGM for GDA, we fine tune the U-Net to map realistic faces to primary avatars using the data generated in Section~\ref{sec:dataset}. The training process is shown in Figure~\ref{fig:domain}, where the realistic face domain evolves into the Bitmoji domain. 
At inference time, we pass a user's reference image $I_\text{ref} \in \mathbb{R}^{3 \times 512 \times 512}$ through the U-Net to map to the pixel-aligned Gaussian parameters of scaling $\boldsymbol{s}$, position $\boldsymbol{t}$, color $\boldsymbol{c}$, opacity $\boldsymbol{o}$, and orientation $\boldsymbol{q}$:
\begin{equation}
\theta _\text{GDA}= \left\{ 
\boldsymbol{s}^k, 
\boldsymbol{t}^k, \boldsymbol{q}^k, \boldsymbol{c}^k, \boldsymbol{o}^k \right\}_{k=1}^M = 
\mathcal{E}_{\text{GDA}}(I_\text{ref}; \Phi_{\text{GDA}}).
\end{equation}
$M$ is the number of Gaussians and $\Phi_{\text{GDA}}$ is the learnable parameters.
The 3D Gaussians are then rendered in the frontal view  $I_\text{sty} = 
\mathcal{E}^\text{render}_\text{3D}(\theta _\text{GDA} )$. 
This process transforms real face photos into the primary 2D avatar domain while preserving identity-related features.

\noindent \textbf{Dual-Stylization.} After GDA, we employ a Stable Diffusion~\cite{Rombach2021HighResolutionIS} pipeline for dual stylization. The diffusion model takes the GDA output image $I_\text{sty}$, a text prompt, and the original user photo as input, and outputs a dual-stylized avatar. First, to preserve the coarse structure of the primary avatar, we use SDEdit~\cite{Meng2021SDEditGI} to start denoising from a noised GDA output. To further preserve the avatar's primary style, we feed the GDA image's Canny edges to ControlNet~\cite{Zhang2023AddingCC}. To maintain identity preservation, we input the original user photo to IP Adapter~\cite{Ye2023IPAdapterTC} to condition the generation on facial similarity embeddings. 
Using the DDIM scheduler~\cite{Song2020DenoisingDI}, the entire process only uses $T = 10$ denoising steps, taking less than one second.

\subsection{2D to 3D Avatar Generation and Animation}
\label{sec:3d_gen}

\noindent \textbf{Expression Encoder.} Current avatar animation techniques using Gaussian Splats often solely depend on 3D Morphable Models (3DMM)~\cite{chu2024gagavatar,hu2024gaussianavatar,qian2023gaussianavatars,shao2024splattingavatar,Song2024TextToonRT}, which limits generalization beyond realistic faces, especially for cartoon avatars. To overcome these constraints, we condition the 3D generation network on not only 3DMM features, but also blendshape weights derived from the Facial Action Coding System (FACS)~\cite{Ekman1978FacialAC}. These weights are widely used in cartoon animation to control facial features like eye position and mouth shape, and can generalize beyond realistic faces.
As depicted in Fig.~\ref{fig:img2avatar}, for generating expressive avatars, we extract expression codes \( f_\text{mm} \in \mathbb{R}^{100} \) from the driving image using a 3DMM estimator. These codes are concatenated with the blendshape vector \( f_\text{bs} \in \mathbb{R}^{16} \), producing a comprehensive expression feature. A learnable projection layer $\mathcal E_\text{proj}$ then projects this combined feature into a 16-dimensional expression vector $f_\text{exp} = \mathcal E_\text{proj}([f_\text{bs}; f_\text{mm}])$, where $[\cdot]$ indicates feature concatenation. To integrate expressiveness with identity, the driving signal is formulated as:
\begin{equation}
f_\text{drive} = (\mathcal E_\text{mlp}([f_\text{exp}, f_\text{id}]), f_\text{pos}).
\end{equation}
Here, \( f_\text{id} \) is the global identity feature extracted from a reference image \( I_r \) via a frozen DINOv2 backbone~\cite{oquab2023dinov2}, and \( f_\text{pos} \) denotes the position map from 3DMM vertices.

\noindent \textbf{3D Generation Network $\mathcal{G}(\cdot)$.} Given the generated unposed avatars $I_\text{unposed}$ and driving features $f_\text{drive}$ from the expression encoder, we employ an asymmetric U-Net architecture akin to Large Multi-view Gaussian Models~\cite{Tang2024LGMLM} and incorporate cross-attention layers to merge the driving features seamlessly:
\begin{equation}
    I_\text{posed} = \mathcal{E}_\text{2D}^\text{render} (\mathcal{G}(I_\text{unposed}, f_\text{drive}; \Phi_g)),
\end{equation}
where $\Phi_g$ is the learnable parameters of $\mathcal{G}(\cdot)$ and $\mathcal{E}_\text{2D}^\text{render}$ is a 2DGS renderer~\cite{Huang2DGS2024}.
$\mathcal{G}(\cdot)$ consists of an encoder with five down-sampling blocks, a middle block, and a decoder with three up-sampling blocks. 
Cross-attention modules are strategically placed in the deeper layers of the network: the last two down-sampling blocks, the middle block, and the first two up-sampling blocks.

\noindent \textbf{Mobile AR Application.} Our framework is designed to facilitate real-time animation on mobile devices. Offline, we use the 3D Generation Network $\mathcal{G}(\cdot)$ from the pipeline shown in Fig.~\ref{fig:img2avatar} to initially generate a base set of Gaussians for the avatar in a rest pose $\theta_{\text{rest}}$, along with specific Gaussian sets corresponding to each component of the expression features $f_\text{drive}$. 
On the mobile device, we use a face tracker, such as \textit{Mediapipe}'s BlazeFace tracker~\cite{Bazarevsky2019BlazeFaceSN}, to generate a list of blendshape weights $f_{\text{bs}} \in \mathbb{R}^{16}$. We leverage these weights to animate the avatar through linear interpolation between the parameters of each feature component:
\begin{equation}
\theta_\text{mobile} = \theta_\text{rest} + \sum_{i=1}^{K} f_{\text{drive}} ^i (\theta_i - \theta_{\text{rest}})
\end{equation} 
$K$ represents the number of driving features, and can be tuned to balance expression detail and speed. For compatibility with Mediapipe, we choose $K=16$.
The final rendering of the Gaussians $\theta_\text{mobile}$ takes place in WebGL, offering efficient rendering while retaining high visual fidelity. To demonstrate this capability, we developed a JavaScript application that allows users to control their avatars directly in their browsers.

\subsection{Training and Losses}
\label{sec:training_losses}

\noindent \textbf{Image to Avatar Generation.} Our training process uses GDA to map the reference identity $I_\text{ref}$ to Gaussian parameters $\theta_\text{GDA}$, which are then used to render the unposed primary avatar $I_\text{unposed}$ from the frontal view via a 3DGS renderer $\mathcal{E}_\text{3D}^\text{render}$. The rendered image is supervised using a combination of Mean Squared Error (MSE) and perceptual LPIPS~\cite{Zhang2018TheUE} losses:

\begin{equation}
    \mathcal{L}_\text{GDA} = \mathcal{L}_{\text{MSE}}(I_{\text{ref}}, I_{\text{sty}}) + \mathcal{L}_{\text{LPIPS}}(I_{\text{ref}}, I_{\text{sty}}).
\end{equation}
 Despite potential noise introduced by low-quality GAN inversion, the pre-training on 3D datasets including Objaverse equips our network with strong generalization capabilities, enabling effective real-to-avatar domain adaptation.
As shown in Fig.~\ref{fig:domain}, GDA efficiently transforms realistic faces into a primary style while preserving the subjects' identity and enhancing features, \textit{e.g.}, eye size.

\noindent \textbf{3D Animatable Stylized Avatar Generation.} To improve the surface geometry of avatars, our model incorporates normal consistency and depth distortion losses. The normal consistency loss $\mathcal{L}_{\text{normal}}$ aligns the normals of 2D Gaussians~\cite{Huang2DGS2024} with surface normals determined through finite differences from rendered depths, thereby reducing noise. Meanwhile, the depth distortion loss $\mathcal{L}_{\text{dist}}$, implemented following \cite{barron2021mip,barron2022mip}, encourages Gaussians to cluster closely along camera rays, effectively enhancing surface representation. This optimization allows our network to output avatars with detailed geometry, suitable for applications such as animation and relighting. 
The total loss function for the 3D generation network is defined as:
\begin{equation}
    \mathcal{L}_{\text{3DGen}} = \mathcal{L}_{\text{render}} + \lambda_{\text{LPIPS}}\mathcal{L}_{\text{LPIPS}} + \lambda_{\text{n}}\mathcal{L}_{\text{normal}} + \lambda_{\text{d}}\mathcal{L}_{\text{dist}},
\end{equation}
where $\mathcal{L}_{\text{render}}$ combines RGB and alpha mask losses:
\begin{equation}
    \mathcal{L}_{\text{render}} = \|{I}_{\text{posed}} -{I}^\text{gt}_\text{posed}\|_2^2 + \|\mathbf{\alpha}^{\text{pred}} - \mathbf{M}^{\text{gt}}\|_2^2.
\end{equation}
$\mathcal{L}_{\text{normal}}$ aligns predicted normals with surface normals:
\begin{equation}
    \mathcal{L}_{\text{normal}} = 1 - (\mathbf{n}^{\text{pred}} \cdot \mathbf{n}^{\text{surf}}).
\end{equation}
Here, ${I}^{\text{posed}}, {I}^\text{posed}_{\text{gt}}$ are the predicted and ground truth images; $\mathbf{\alpha}^{\text{pred}}$ and $\mathbf{M}^{\text{gt}}$ are the predicted alpha mask and its ground truth counterpart; $\mathbf{n}^{\text{pred}}, \mathbf{n}^{\text{surf}}$ are predicted and surface normal vectors. 
$\lambda_{\text{lpips}}, \lambda_{\text{n}}, \lambda_{\text{d}}$ are weights for $\mathcal{L}_{\text{lpips}}$, $\mathcal{L}_{\text{normal}}$ and $\mathcal{L}_{\text{dist}}$, respectively. The normal and distortion losses commence after 20\% of training to first establish basic appearance convergence.

\section{Experiments}
As no prior work follows the exact design goals as ours, we evaluate each component of the \methodname{} system individually, similar to an ablation study. Section~\ref{subsec:exp_img2img} studies the image to 2D avatar generation step, Section~\ref{subsec:2dstyle23d} studies the 2D-to-3D avatar generation step, and Section~\ref{subsec:animation} studies the 3D animation step. Evaluation is performed on a dataset of Bitmojis which we plan on open sourcing upon publication. 

\begin{figure*}[t]
\centering
\includegraphics[width=0.9\textwidth]{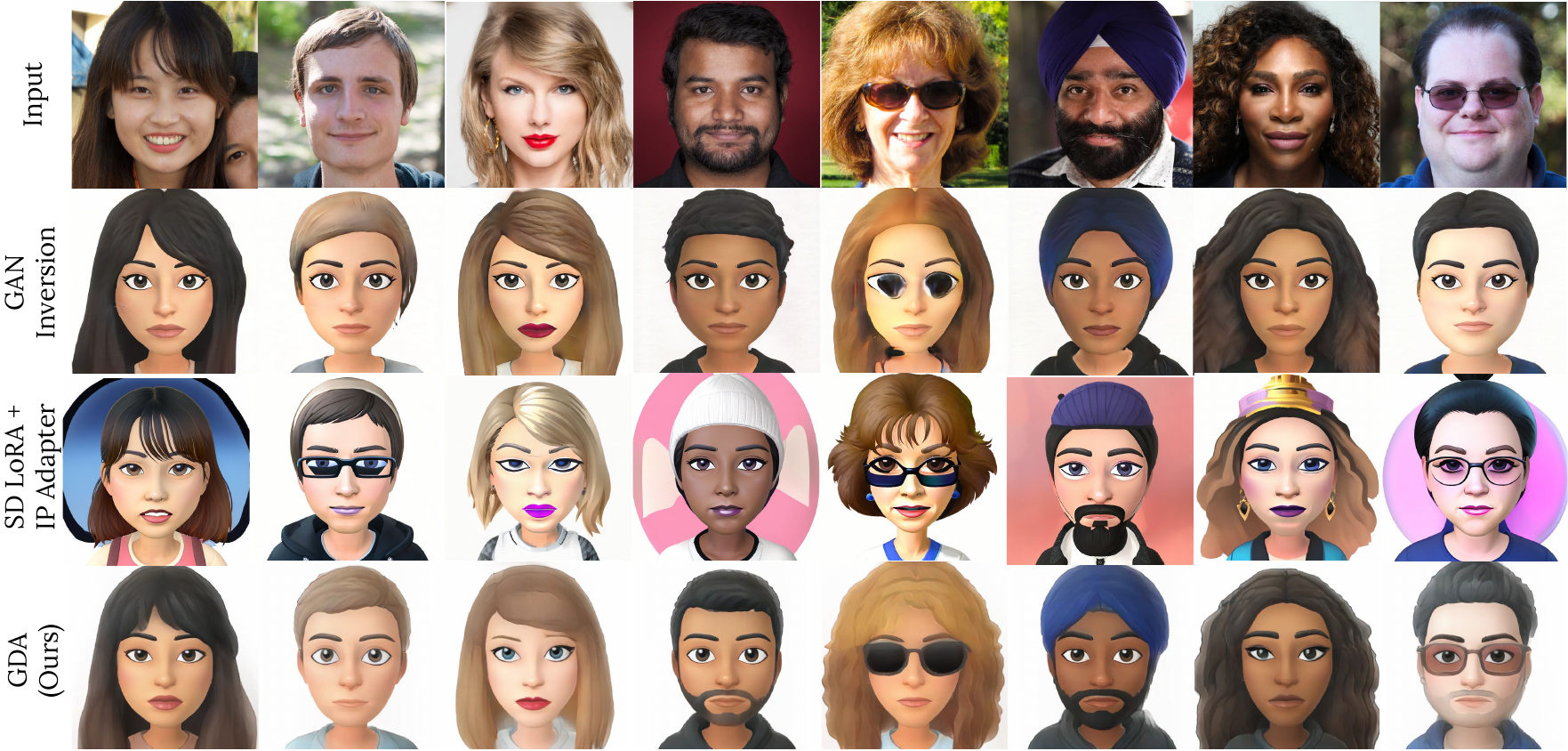}
\caption{\textbf{Image to 2D Avatar Generation.} Showcased are photos from eight individuals transformed into the primary Bitmoji domain using various methods. GAN inversion produces overly generic avatars, struggling with unique features such as beards, glasses, and headwear. Diffusion-based models inaccurately add features, making them inconsistent for targeted styles. In contrast, our GDA method excels in creating high-quality avatars, effectively retaining the original identity features.}

\label{fig:gda_results}
\end{figure*}

\begin{figure*}[t]
\centering
\includegraphics[width=0.9\textwidth]{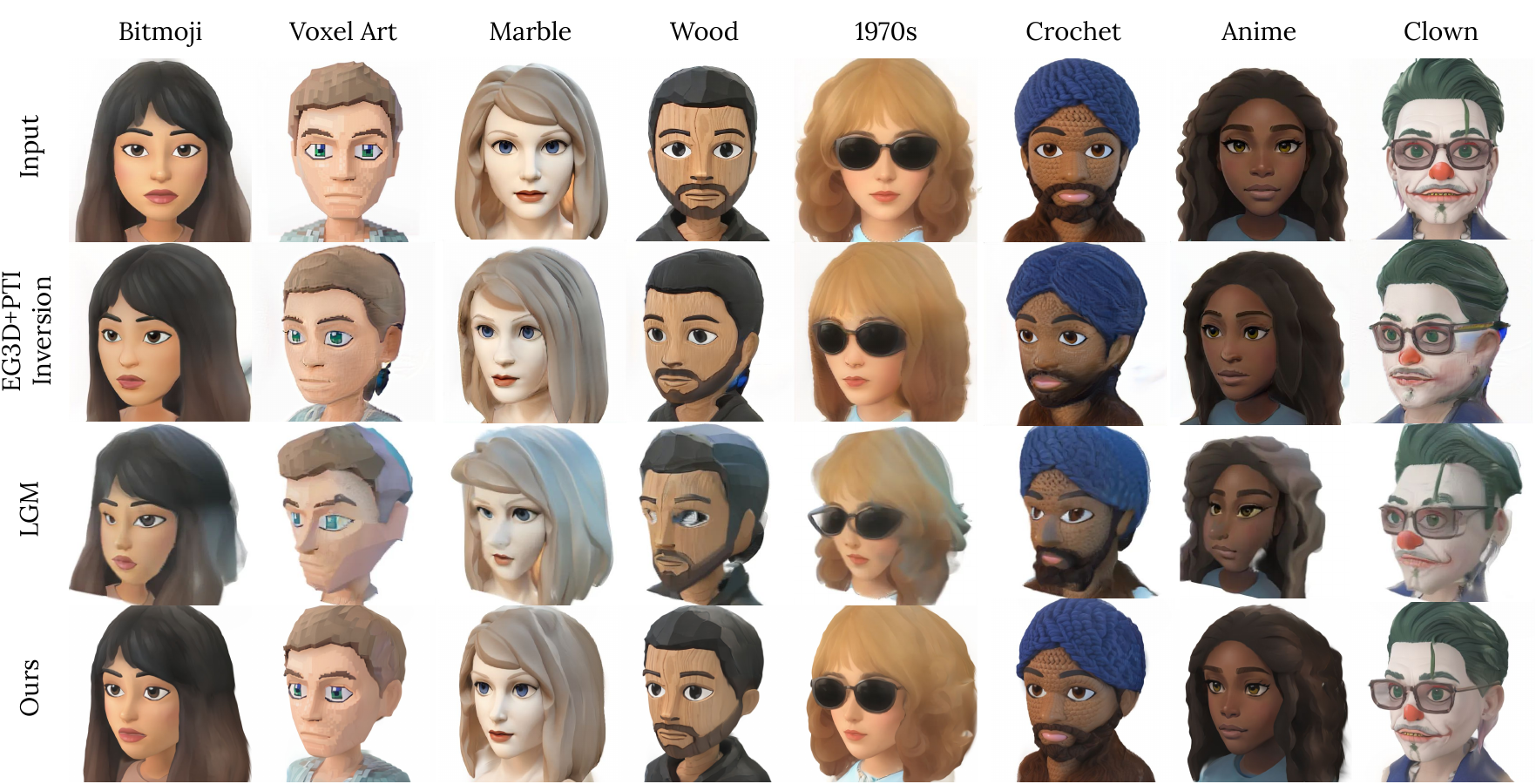}
\caption{\textbf{2D to 3D Avatar Generation.} We demonstrate the process of converting dual-stylized avatar, derived from the primary avatars in Fig.~\ref{fig:gda_results}, into 3D avatars. PTI inversion with EG3D~\cite{Chan2022,Roich2021PivotalTF} struggles to accurately reproduce 3D geometry, while LGM~\cite{Tang2024LGMLM} produces artifacts in both geometry and texture. Despite being trained exclusively on the Bitmoji style, our method successfully generates high-quality 3D avatars in previously unseen styles.}
\label{fig:3d_comparison}
\end{figure*}

\begin{figure*}[ht]
\centering
\includegraphics[width=0.9\linewidth]{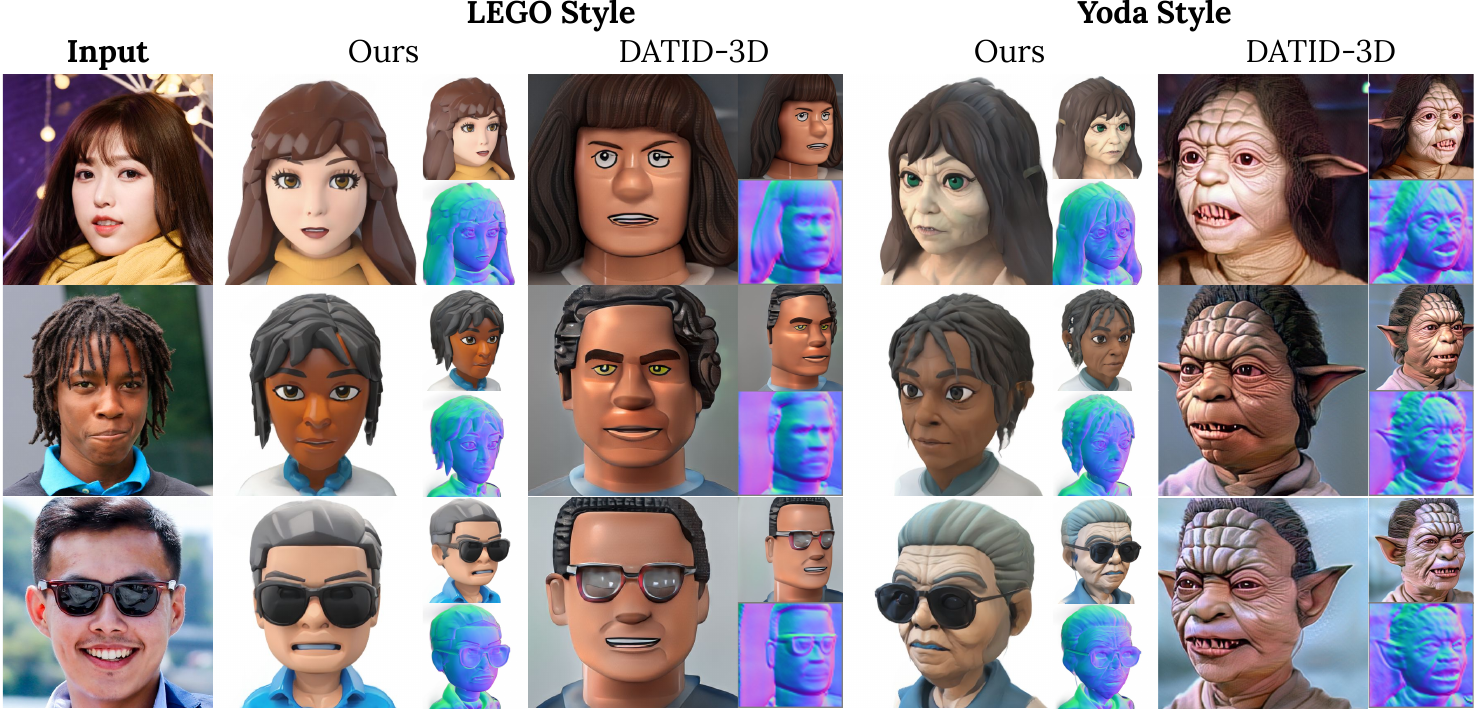}
\caption{\textbf{Single Portrait to 3D Generation.} We compare \methodname{} with DATID-3D~\cite{kim2022datid3d} in the context of dual stylization. For each method and style, we render outputs from two viewpoints alongside a normal map. DATID-3D exhibits typical GAN-related issues, such as poor identity preservation and limited stylistic diversity, resulting in similar outputs across different identities. Conversely, \methodname{} effectively maintains identity and produces distinct styles, showcasing superior image quality and sharper geometry.}
\label{fig:3dtoon}
\end{figure*}

\subsection{Image to 2D Avatar Generation}
\label{subsec:exp_img2img}

\begin{table}[t]
\centering
\renewcommand\tabcolsep{15pt}
\caption{\textbf{Image to 2D Avatar Generation.} We compare different methods of generating 2D stylized avatars. Our GDA significantly outperforms GAN inversion and diffusion in terms of image quality (FID, KID), identity preservation (ID), and execution speed.}
\resizebox{1\linewidth}{!}{
\begin{tabular}{c|c|c|c|c}
    \toprule
     & FID $\downarrow$ & KID $\downarrow$ & ID $\uparrow$ & Speed $\downarrow$ \\
     \midrule
     GAN Inversion & 93.73 & 0.0603 & 0.16 & 98.14s\\
     Diffusion & 93.63 & 0.0457 & 0.19 & 3.54s \\
     GDA (Ours) &  \textbf{72.94}&\textbf{0.0346} & \textbf{0.25} & \textbf{0.080s} \\
     \bottomrule
\end{tabular}
}

\vspace{-10pt}
\label{tab:gda_metrics}
\end{table}

\noindent \textbf{Baselines.} We first evaluate GDA for image-to-avatar generation and compare with GAN inversion and diffusion-based methods, both fine-tuned on our Bitmoji dataset. For GAN inversion, invert real faces into the latent space of a fine-tuned SemanticStyleGAN \cite{Shi2021SemanticStyleGANLC} model. The diffusion baseline is a Stable Diffusion 1.5 model \cite{Rombach2021HighResolutionIS} fine-tuned with a Bitmoji LoRA \cite{Hu2021LoRALA} and BLIP-2 \cite{Li2023BLIP2BL} captions. At inference time, a real face is input into an IP Adapter Plus Face~\cite{Ye2023IPAdapterTC} model for identity conditioning.

\noindent\textbf{Evaluation.} We conducted an evaluation using 100 randomly selected faces from the FFHQ dataset, assessing each method on visual quality, identity retention, and speed. Visual quality was measured through FID and KID scores between the transformed images to the Bitmoji dataset. Identity retention was evaluated using ArcFace~\cite{deng2019arcface}, and speed was benchmarked on a Nvidia L4 GPU.
As reported in Table \ref{tab:gda_metrics}, GDA outperforms the baselines across all metrics, achieving FID scores more than 20 points lower than those of GAN inversion and diffusion. Figure~\ref{fig:gda_results} visually highlights these quality differences: GAN inversion GAN inversion struggles to generate avatars with diverse hair styles, eye colors, and clothing. The diffusion approach fails to maintain a consistent style and often incorrectly introduces features like glasses, undermining both style and identity preservation.
In contrast, GDA produces avatars that both have a consistent Bitmoji style and retain key identity features such as eye color, sunglasses, and hairstyles. Finally, since GDA only requires one forward pass through a UNet, it is two and four orders of magnitude faster than diffusion and GAN inversion respectively, translating images in less than 0.1 seconds.  Surprisingly, even though GDA is trained on data generated from GAN inversion, because of GDA's Objaverse~\cite{Deitke2022ObjaverseAU}  prior, the resulting images are more detailed than from GAN inversion alone.

\subsection{3D Avatar Generation }
\label{subsec:2dstyle23d}

\noindent\textbf{2D to 3D Avatar Generation.} After dual stylization, our method lifts the 2D avatar image to 3D. We compare our proposed technique against two other single-image 3D reconstruction techniques: a EG3D \cite{Chan2022} model fine-tuned on Bitmojis, and LGM \cite{Tang2024LGMLM}. The EG3D baseline maps 2D images to radiance fields via pivotal tuning inversion~\cite{Roich2021PivotalTF}, while LGM uses MVDream~\cite{shi2023MVDream} to transform a single image into multi-view images, then maps that set of images to 3D Gaussians. 
We assessed each method using 100 random 3D Bitmojis, rendered from ten views distributed spherically around the head. As shown in Table \ref{tab:3d_metrics}, our method surpasses all baselines, demonstrating superior capability in accurately converting 2D images to 3D, while being significantly faster, needing only a single U-Net pass.

Figure~\ref{fig:3d_comparison} provides visual comparisons on dual-stylized avatars. The top row features eight dual-stylized avatars generated  from the identities in Fig.~\ref{fig:gda_results}. EG3D struggles to generate high-fidelity geometry. Similarly, due to the diffusion process in MVDream, LGM is a slow, iterative process, and often produces incorrect 3D head geometries. In contrast, our method successfully creates high-quality textures and geometry, even for out-of-distribution accessories like turbans and sunglasses.

\begin{table}[t]
\centering
\renewcommand\tabcolsep{10pt}
\caption{\textbf{2D to 3D Avatar Generation.} Our approach outperforms EG3D~\cite{Chan2022} and LGM~\cite{Tang2024LGMLM} on all metrics, providing superior texture and geometry accuracy with faster processing.}
\resizebox{1\linewidth}{!}{
\begin{tabular}{c|c|c|c|c}
    \toprule
     & PSNR $\uparrow$ & SSIM $\uparrow$ & LPIPS $\downarrow$ & Speed $\downarrow$ \\
     \midrule
     EG3D~\cite{Chan2022} & 10.92 & 0.68 & 0.50 & 95.1s\\
     LGM~\cite{Tang2024LGMLM} & 12.16 & 0.69 & 0.53 & 2.82s \\
     Ours & \textbf{18.73}  & \textbf{0.81} & \textbf{0.24} & \textbf{0.091s}\\
     \bottomrule
\end{tabular}
}

\vspace{-5pt}
\label{tab:3d_metrics}
\end{table}

\noindent\textbf{Single Portrait to 3D Generation.}
Next, we evaluate \methodname{}'s full image-to-3D-avatar generation ability against DATID-3D~\cite{kim2022datid3d}. Figure ~\ref{fig:3dtoon} shows that DATID-3D struggles to maintain the original identity in avatars. 
\methodname{}, however, achieves a robust balance of identity preservation and style versatility. Our approach produces sharp images with detailed geometries, and only creates each avatar in just 0.9 seconds, a significant improvement over DATID-3D's 90-second processing time. 

\noindent\textbf{Dual Stylization User Study.} Recall that one of our major goals is to support dual stylization. The avatars should resemble the user, fit the art style of the Bitmoji world, and enable diversity in customization. We compare \methodname{} to DATID-3D with a user study on the 12 avatars in Figure~\ref{fig:3dtoon}. Across 27 participants, 92\% said that our avatars better resemble the input users. The DATID-3D avatars are not diverse enough to preserve the appearance and skin tone of users, making them unacceptable for a production system. Regarding style, 96\% said our avatars had more variation in style and identity, and 96\% said they would prefer our avatar aesthetics in a video game set in the Bitmoji world. While the avatars generated from DATID-3D may look more like \lego{} or \yoda{} at first glance, they are limited in diversity, and do not fit the primary Bitmoji style. 
\begin{table}[t]
\centering
\renewcommand\tabcolsep{4pt}
\caption{\textbf{User Preferences for Dual-Stylized Avatars.} Our method outperforms DATID-3D~\cite{kim2022datid3d} in user preferences. Users believed our avatars better preserved idenity, had more diversity, and would be a better fit the aesthetics of a Bitmoji video game. }
\resizebox{1\linewidth}{!}{
\begin{tabular}{c|c|c}
    \toprule
     User Study (N=27)&   DATID-3D & Ours\\
     \midrule
     Better identity preservation? &8\% & 92\% \\
     Better variation in style and identity? & 4\% & 96\% \\
     Better aesthetics in a Bitmoji video game?  & 4\% & 96\% \\
     \bottomrule
\end{tabular}
}
\vspace{-10pt}

\label{tab:user_study}
\end{table}

\subsection{3D Avatar Animation } 
\label{subsec:animation}

\begin{figure}[t]
\includegraphics[width=1\linewidth]{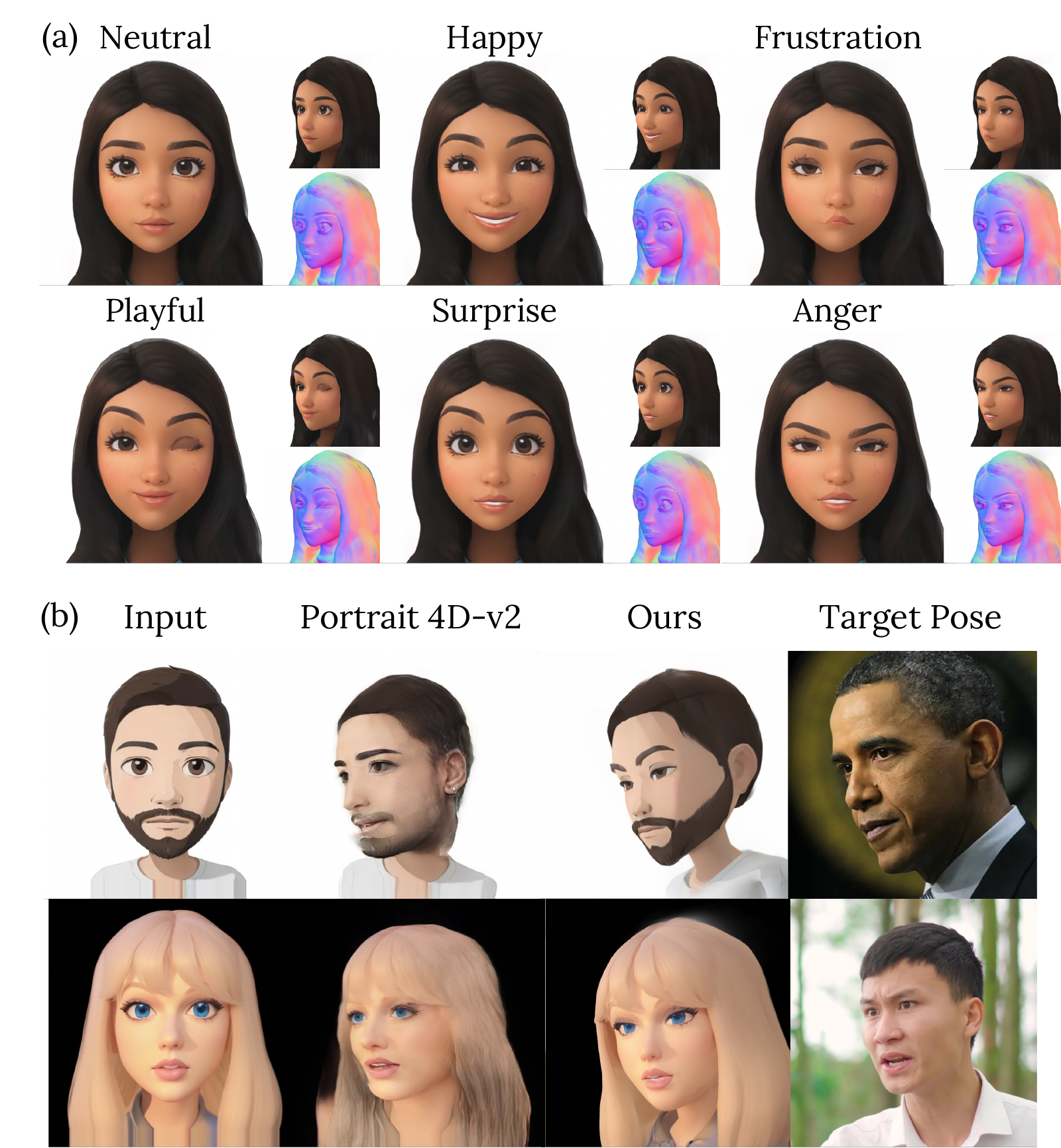}
\caption{\textbf{Avatar Animation.} (a) An \methodname{} showcasing various emotions controlled by blendshape weights. (b) \methodname{} effectively transfers expressions from driving images, outperforming Portrait4D-v2~\cite{deng2024portrait4d} in accuracy and visual appeal.} 
\label{fig:animation_sum}
\end{figure}

\noindent \textbf{Expression Animation.} \methodname{} enables avatars to express a wide range of emotions, such as neutrality, happiness, frustration, playfulness, anger, and surprise, by using blendshape weights, as shown in Fig.~\ref{fig:animation_sum}(a). Additionally, \methodname{} can perform expression transfer from driving images, producing 3D-consistent and visually appealing avatars. Fig.~\ref{fig:animation_sum}(b) shows this capability, where \methodname{} outperforms Portrait4D-v2~\cite{deng2024portrait4d} by generating avatars with more accurate expressions derived from the target image. 3DMMs, used by  Portrait3D-v2~\cite{deng2019arcface} and many other works~\cite{Song2024TextToonRT,he2025lam,chu2024gagavatar}, do not generalize to the geometry of cartoon heads.

\begin{figure}[t]
\includegraphics[width=1\linewidth]{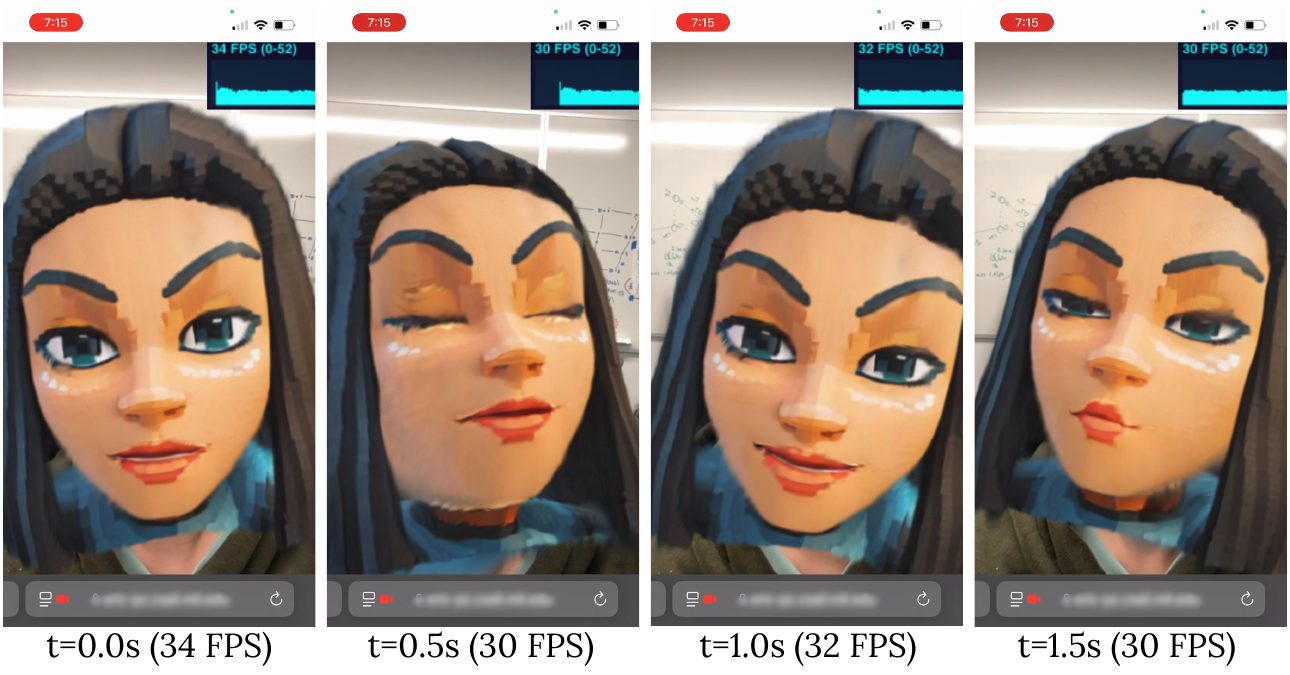}
\caption{\textbf{Mobile AR.} \methodname{} enables a user to puppet their avatar in augmented reality. Our method runs entirely on a web client, at 30-40 FPS on a phone. 
}
\vspace{-10pt}
\label{fig:mobile}
\end{figure}

\noindent\textbf{Mobile AR.}  
\label{subsec:mobile}
We provide a web-based AR app to efficiently render and animate avatars using a face tracker.  As illustrated in Fig.~\ref{fig:mobile}, an avatar animated using a user's facial expressions can be rendered at 30--40 FPS on an iPhone 13 Pro. These avatars occupy only 3 MB of disk space, enabling the creation of dynamic filters and engaging AR effects directly within a mobile web browser. 
To highlight the advantages of our animation technique, we compare \methodname{} against TextToon~\cite{Song2024TextToonRT}. Like many other avatar generation methods~\cite{hu2024gaussianavatar, qian2023gaussianavatars, chu2024gagavatar}, TextToon requires a neural network to predict 3DMM features, limiting it to 15–18 FPS on an M1 MacBook. In contrast, our method consistently runs at 90-100 FPS on a laptop. Moreover, TextToon's dependence on 3DMMs limits its practicality on phones, whereas our cross-platform solution can still run at 30 FPS. Table \ref{tab:rendering_performance} offers a detailed feature comparison. Other work like LAM~\cite{he2025lam} propose alternative ways to render avatars on mobile devices, but we do not compare against them because they are incapable of AR puppeting. Unlike our model and TextToon, they do not attempt to integrate their system with a face tracker at inference time.

\begin{table}[t]
\centering
\renewcommand\tabcolsep{2pt} %
\caption{\textbf{Mobile AR Application Comparison.} We compare various features of our mobile AR application and TextToon~\cite{Song2024TextToonRT}.}
\resizebox{1\linewidth}{!}{
\begin{tabular}{c|cc|cc}
    \toprule
    \multirow{2}{*}{\centering Method} & \multicolumn{2}{c|}{Frame Rate (FPS)} & \multirow{2}{*}{Cross-Platform} & \multirow{2}{*}{Driving Signal} \\
           & M1 MacBook & iPhone 13 Pro &  & \\
    \midrule
    TextToon~\cite{Song2024TextToonRT} & 15--18 & N/A & \texttimes & 3DMM \\
    \methodname{} (Ours) & 90-100 & 30--40 & \checkmark & 3DMM + Blendshapes \\
    \bottomrule
\end{tabular}
}

\label{tab:rendering_performance}
\vspace{-5pt}
\end{table}

\section{Conclusion}

Although avatar generation is now one of the most popular research areas in computer vision, the gap between the existing research literature and production is still large. We aim to bridge this gap by introducing \methodname, a system for generating animatable, dual-stylized avatars from selfies almost instantly. Leveraging Gaussian Domain Adaptation, \methodname{} first converts selfies into primary stylized avatars, then applies a diffusion process for a secondary style while preserving identity integrity. The system achieves selfie-to-avatar conversion in just 0.9 seconds, and enabling real-time interactions at 30--40 FPS. 

\noindent\textbf{Limitations and Future Work.} \methodname~requires a large 3D avatar dataset for training. We perform all experiments on Bitmoji avatars, but future work could also apply these methods to other avatar platforms. We recognize that there is a lack of avatar data available for research, so we intend to publicly release our data as well. Future improvements could include improving the face tracker, which limits the quality of the animations. Mediapipe can produce noisy blendshapes, leading to jittery results.

{
    \small
    \bibliographystyle{ieeenat_fullname}
    \bibliography{sample-bibliography}
}
 \appendix
 \clearpage
\setcounter{page}{1}

\appendix
\section{Implementation Details}

\begin{figure}[h]
\centering
\includegraphics[width=\linewidth]{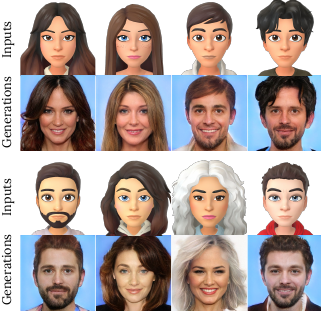}
\caption{\textbf{GDA Training Data.} Visualization of data pairs used to train the GDA network. Each avatar is inverted into the latent space of a GAN, and a generator trained on realistic faces creates a corresponding realistic face, preserving features such as hairstyles, hair colors, and general facial characteristics.}
\label{fig:inv_data}
\end{figure}

\subsection{Bitmoji Training Data}
\paragraph{GDA Training Data.} To train our Gaussian Domain Adaptation network, we start with a dataset of random Bitmojis and use GAN inversion to generate corresponding realistic images. Fig.~\ref{fig:inv_data} showcases some examples of the training data generated through this process. The resulting faces mirror the hairstyles, hair colors, and facial features of the original avatars. While the generated images may contain artifacts and exhibit limited diversity, our GDA model benefits from pre-training on Objaverse~\cite{Deitke2022ObjaverseAU}, enabling it to leverage prior knowledge and produce more detailed reconstructions than GAN inversion alone. This approach enhances the accuracy and expressiveness of the domain adaptation process.

\begin{figure}[t]
\centering
\includegraphics[width=\linewidth]{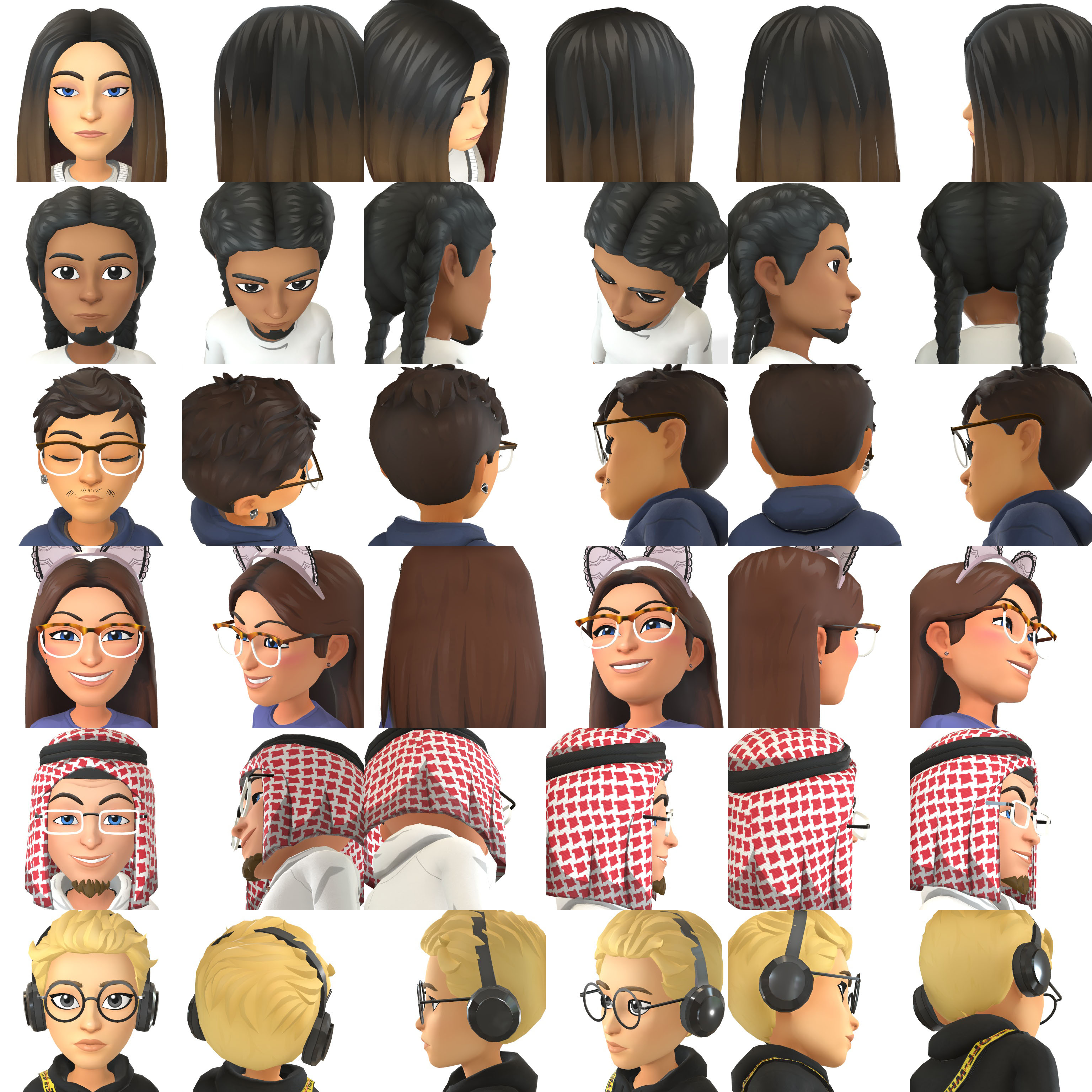}

\caption{\textbf{Multi-view Bitmoji Training Data.} Samples from the Bitmoji dataset used in training the 3D Generation Network. Avatars are rendered from the front and multiple random angles around the head, with random blendshapes applied to simulate various expressions.}
\label{fig:bitmoji_mv}
\end{figure}

\paragraph{Multi-view Training Data.}
Fig.~\ref{fig:bitmoji_mv} visualizes training samples from the Bitmoji dataset used for training our 3D Generation Network. Each avatar is rendered from 10 spherically distributed viewpoints around the head and is posed with random blendshape weights to simulate diverse facial expressions. The dataset features a wide range of hairstyles, skin tones, and accessories such as glasses, hats, and earrings. Although the U-Net is trained exclusively on Bitmoji-style avatars, it effectively reconstructs dual-stylized avatars that exhibit distinct appearances and textures, demonstrating the network's versatility and generalization capability.

\subsection{Facial Action Coding System}
We implement the following 16 blendshapes from the Facial Action Coding System. These blendshapes are compatible with most facial blendshape predictors like \textit{Apple ARKit} or \textit{Google Mediapipe}. 

\begin{itemize}
    \item \texttt{browDownLeft}
    \item \texttt{browDownRight}
    \item \texttt{browUpLeft}
    \item \texttt{browUpRight}
    \item \texttt{eyeBlinkLeft}
    \item \texttt{eyeBlinkRight}
    \item \texttt{jawOpen}
    \item \texttt{jawLeft}
    \item \texttt{jawRight}
    \item \texttt{lipsPucker}
    \item \texttt{mouthFrownLeft}
    \item \texttt{mouthFrownRight}
    \item \texttt{mouthSmileLeft}
    \item \texttt{MouthSmileRight}
    \item \texttt{mouthStretchLeft}
    \item \texttt{mouthStretchRight}
\end{itemize}
\label{app:facs}

To animate the Bitmoji avatars for training data, we used the publicly available Bitmoji rig available here: \href{https://developers.snap.com/lens-studio/features/bitmoji-avatar/animating-bitmoji-3d}{[https]}.

At inference time, we can use a real-time blendshape predictor like \textit{ARKit} or \textit{Mediapipe} to puppet the avatars from a real video. Please see the attached HTML gallery for a demonstration. 

\subsection{User Interfaces}

To showcase the intuitive features of the \methodname{} system, we present videos of the interface interactions available in the HTML gallery. The avatar generation interface, crafted with \textit{Gradio}, enables users to effortlessly create dual-stylized avatars from their own photos. In addition, the blendshape editor, developed using \textit{Viser}, allows users to pose their avatars in 3D by adjusting blendshape weights, thereby controlling facial expressions.

During the diffusion stylization process, we provide users with key parameters to balance identity preservation with style diversity. These controls are listed by their significance:
1) Style Transition Strength; 2) Edge Preservation Level 3) Identity Consistency Factor.

\noindent \textit{Style Transition Strength}: This parameter, inspired by methods similar to SDEdit, regulates the extent of the stylization transition. Lower values enable the dual-stylized avatar to retain more details from the original single-styled avatar input.

\noindent \textit{Edge Preservation Level}: This setting influences how accurately the system maintains the structure of the avatar by preserving the edges from the single-styled input.

\noindent \textit{Identity Consistency Factor}: This controls the strength of identity features from the initial input photo, ensuring that essential facial characteristics remain recognizable.

We encourage users to view the videos in the HTML gallery to observe how these parameters affect avatar generation, enhancing both creativity and user experience.
\section{Additional Results}

\subsection{Results Gallery}

We invite you to explore the HTML gallery, which features videos of \methodname avatars animated in 3D. Access the gallery by opening the \texttt{index.html} file in your web browser. The gallery includes the following highlights:

\begin{enumerate}
    \item Dual-stylized avatars with dynamic facial animations displayed from various novel viewpoints.
    \item A demonstration of the avatars' capabilities in facial puppeting for augmented reality applications.
    \item Screen captures of the \methodname user interfaces, showcasing the ease of creating dual-stylized avatars and posing them using blendshapes.
\end{enumerate}

\subsection{More Applications}

\begin{figure}[t]
\centering
\includegraphics[width=\linewidth]{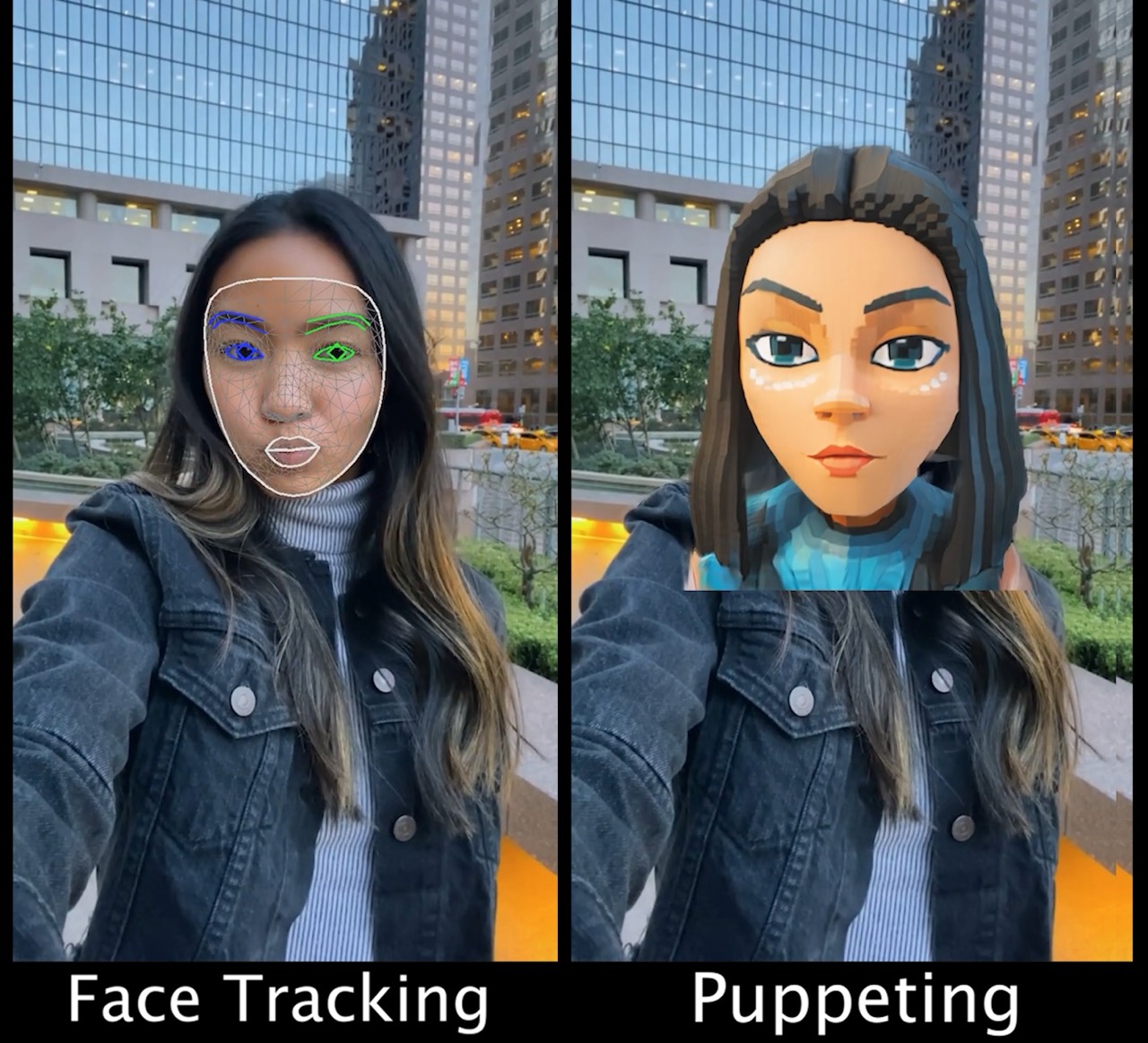}
\caption{\textbf{Augmented Reality Puppeting.} This example demonstrates the use of \textit{Mediapipe}'s real-time face detection to animate avatars based on estimated blendshape weights. By alpha-compositing the avatars with the original input, we enable dynamic puppeting in augmented reality. For live demonstrations, please refer to the HTML gallery.}
\label{fig:ar_avatar}
\end{figure}

\noindent\textbf{3D Avatar Animation.} Dual-stylization offers the ability to swiftly visualize avatars in various scenarios, unlocking numerous applications. As illustrated in Fig.~\ref{fig:teaser}, \methodname{} avatars can be employed to create personalized comics and stickers, offering users a unique way to express themselves. Another promising application lies in augmented reality (AR), where avatars can be controlled and animated with real-time tracked facial expressions. Examples of this application are shown in Fig.~\ref{fig:ar_avatar} and within the HTML gallery. By utilizing \textit{Mediapipe}'s real-time blendshape tracker, we animate the 3D avatars and seamlessly integrate them with video content, enabling them to be rendered in an AR environment through alpha compositing.

\noindent\textbf{Real-time Web Rendering.} Our choice to represent avatars using Gaussian Splats enables efficient real-time rendering on mobile devices. As demonstrated in Fig.~\ref{fig:web} and     in the HTML gallery, the avatars achieve a rendering rate of 90-100 FPS on a laptop, and 30-40 FPS on a phone. When paired with a face tracker, these avatars can be used to generate engaging filters and augmented reality effects. The demonstration showcases an avatar rendered in Google Chrome on a MacBook, entirely on the client side.

\begin{figure}[t]

\includegraphics[width=\linewidth]{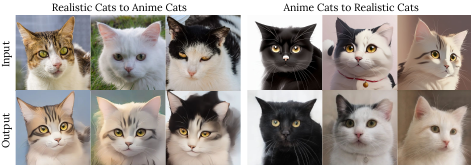}

\caption{\textbf{GDA Generalization Across Domains.} This illustration showcases the versatility of Gaussian Domain Adaptation (GDA) as an image-to-image translation method. Demonstrated above is GDA's capability to transform realistic cat images into anime-style representations and vice versa, highlighting its potential for a wide range of applications beyond avatar creation.}
\label{fig:cats}
\end{figure}

\noindent\textbf{GDA Generalization.} GDA demonstrates that the features learned from few-shot 3D reconstruction models are transferrable to new tasks. Shown in Fig.~\ref{fig:cats}, GDA can be applied for more domains such as cats. We hope that GDA can inspire future work on using Gaussian features for other tasks. 

\section{Ablation Studies}
\begin{figure}[t]
\includegraphics[width=1\linewidth]{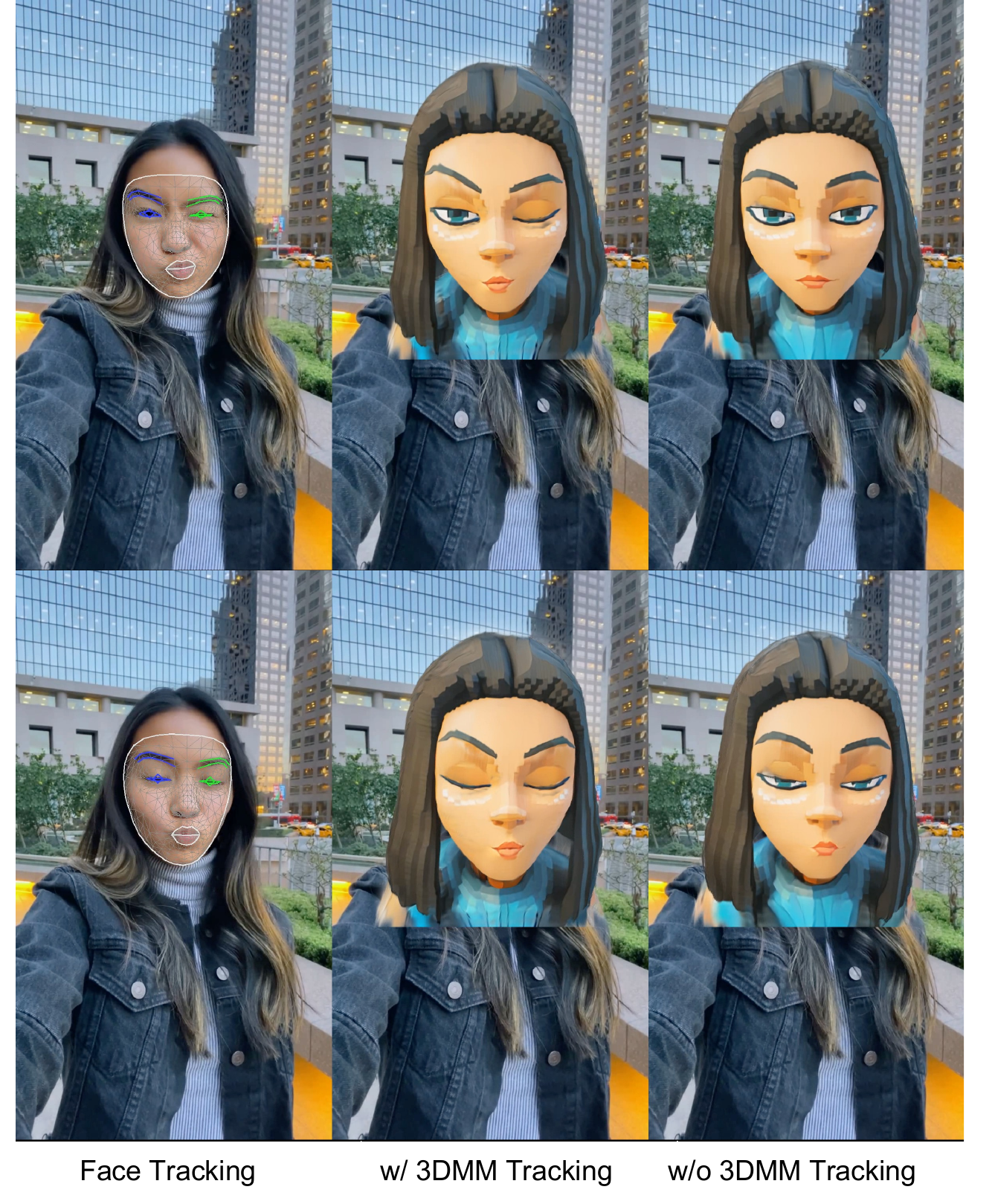}
 \caption{\textbf{Ablation Study on 3DMM Tracking.} This figure demonstrates the effects of using 3DMM features in conjunction with FACS blendshape weights. The combination enhances the expressiveness and fidelity of avatar animation, accommodating both realistic and stylized facial expressions.}

\label{fig:ab_3dmm}
\end{figure}

\subsection{Gaussian Domain Adaptation} 
\begin{figure}[t]
\includegraphics[width=1\linewidth]{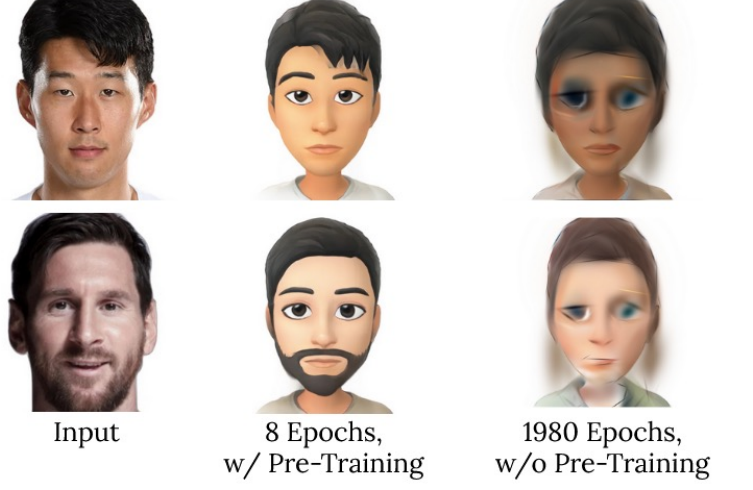}
 \caption{\textbf{Ablation Study on Pre-training.} Objaverse pre-training is necessary to obtain high-quality results from GDA. Without pre-training, the GDA model struggles to converge, even after 250x more epochs than with pre-training.}

\label{fig:ab_pretrain}
\end{figure}

For the task of 2D image-to-image translation, we find that the LGMs, which generate 3D Gaussian splats, are still very well-suited for this task. We believe this is due to two factors: that the LGM model is pre-trained on Objaverse, and that 3D-awareness helps create semantic correspondences in the 2D style transfer task. Shown in Figure~\ref{fig:ab_pretrain}, even after training a LGM model from scratch for 1980 epochs, it still does not converge. In addition, 3D-awareness helps generate accessories like sunglasses, which intuitively should be placed in front of a face. These two priors help the GDA model overcome the quality issues with the data in Fig.~\ref{fig:inv_data}, leading to the high-quality results shown in the main text..

\subsection{3DMM Tracking} As shown in Fig.~\ref{fig:ab_3dmm}, our ablation study highlights the complementary strengths of 3DMM tracking and FACS-based blendshape features in avatar animation. 3DMM is adept at capturing realistic facial expressions, making it ideal for animating real faces, but it struggles with the exaggerated features typical of cartoon avatars. Conversely, FACS blendshapes excel in stylizing facial elements, such as eye and mouth shapes, crucial for cartoon animation. By integrating the precision of 3DMM with the expressive capability of FACS blendshapes, we enhance the overall animation quality, enabling our avatars to faithfully portray both realistic and stylized expressions, thus delivering a more versatile and convincing animation experience.

\section{Ethical Discussion}
The use of photorealistic avatars has raised significant privacy and ethical concerns, particularly in relation to their potential misuse in creating deep fakes and spreading misinformation. In contrast, stylized cartoon avatars offer a safer alternative as they are not easily exploited for direct impersonation. In our work, we have prioritized user privacy by ensuring that no real person’s images are used to train our models. Instead, the realistic images employed for training the Gaussian Domain Adaptation (GDA) system are generated by a GAN. We recognize, however, that GAN-generated data can reflect the biases present in the original datasets used for training. Consequently, we remain vigilant about these limitations and are committed to continuous evaluation and improvement to mitigate any unintended biases.

\end{document}